# Data Envelopment Analysis with Robust and Closest Targets: Integrating Full-Dimensional Efficient Facets for Risk-Resilient Benchmarking

Xiuquan Huang[a], Xi Wang[b], Tao Zhang[b], Xiaocang Xu[c, *], Ali Emrouznejad[d]

[a] *School of Elderly Care Services and Management, Nanjing University of Chinese Medicine, Nanjing 210023, China*

[b] *Faculty of Humanities and Social Sciences, Macao Polytechnic University, Macao 999078, China*

[c] *School of Economics and Management, Huzhou University, Huzhou 313000, China*

[d] *Surrey Business School, University of Surrey, United Kingdom*

Correspondence author: Xiaocang Xu    Email: 03122@zjhu.edu.cn    Tel: 18227553350

Postal address: School of Economics and Management, Huzhou University, Huzhou 313000, China

**Abstract:** As the external environment become increasingly volatile and unpredictable, the selection of benchmarking targets in Data Envelopment Analysis (DEA) should account for their ability to consider risks; however, this aspect has not received sufficient attention. We propose a robust benchmarking target defined by the intersection of the maximum number of full-dimensional efficient facets, each representing a unique marginal substitution relationship. These targets can serve as robust projections for decision making units that are lacking prior risk information because they incorporate the maximum number of marginal substitution relationships. This enables decision makers to adjust their production through these relationships, thereby maximizing the likelihood of achieving globally optimal outcomes. Furthermore, we propose a novel, well-defined efficiency measure based on robust and closest targets. Finally, we demonstrate the application of the proposed measure using a dataset comprising 38 universities from China's 985 Project.

## 1. Introduction

Data Envelopment Analysis (DEA) proposed by Charnes et al. (1978) is a method measuring relative efficiency of a set of decision-making units (DMUs) with several inputs and outputs and has been widely used in many fields of economics and management (Aigner et al., 1977; Emrouznejad et al., 2023; Emrouznejad et al., 2025; Mergoni et al., 2024; Zhu, 2022). In addition to efficiency measurement, DEA has many other abilities, such as providing benchmarking targets and estimation of substitutional rates (Olesen & Petersen, 2003). As business environments become more complex and unpredictable, DMUs are increasingly exposed to risks that can disrupt their operations. When disruptions occur, DMUs attempt to adjust their strategies to mitigate the impact (Arabmaldar et al., 2024). However, due to the unpredictable nature of risks and DMUs' limited awareness and capacity to gather information, they may lack any prior knowledge of risks. In addition, their ability to adapt is often limited by existing operational frameworks due to path dependence (Godin et al., 2019; Guo et al., 2020).

During the COVID-19 pandemic, retailers with hybrid online-offline operations quickly adapted to market shifts by transitioning fully online, while offline-only retailers faced significant challenges due to insufficient digital infrastructure and experience. This contrast highlights the importance of a firm's operational framework in managing unexpected risks. When setting benchmarking targets for DMUs, it is essential to balance efficiency with resilience, ensuring targets remain both practical and robust in uncertain environments. This example underscores the inherent risk-resilience of hybrid models, aligning with the principles of Real Options Theory in risk management (Huang et al., 2022). As a result, dual-focus efficiency targets better reflect real-world adaptability.

In this study, we introduce a kind of robust benchmarking target or risk mitigation strategy. In this context, the "robust" refers to a projection target's ability to withstand risks effectively. According to Real Options Theory, DMUs facing uncertainty can mitigate risks and enhance decision-making flexibility by preserving future options. Such projection targets or positions offer multiple candidate production modes, enabling DMUs without prior knowledge of risk information to adapt through marginal substitution. Once risks materialize, DMUs can then select the optimal production mode. Leveraging the property that each full-dimensional efficient facet (FDEF) embodies a unique marginal substitution relationship (MSR), we propose and prove that the intersection of the maximum number of FDEFs represents the robust solution for DMUs operating under uncertainty. In addition, we also require that the projection target satisfies the least distance from the inefficient DMU to the production frontier. Hence, a well-defined efficiency measure is developed based on targets that are both robust and closest to the reference. It is the first to integrate risk management, least distance, and well-defined rates of substitution simultaneously. Additionally, this strategy is based on FDEF theory proposed by Olesen and Petersen (1996), so this study also represents a theoretical extension and application of this theory in the field of risk management. Finally, we make use of one empirical data to demonstrate the applicability of the model in real life.

The rest of the paper is organized as follows. Section 2 presents the literature review. Section 3 introduces the key notation FDEF along with its specific characteristics inherent to this theoretical framework. Section 4 presents a risk-averse Single-FDEF strategy and a multi-FDEF intersection strategy. Section 5 proposes an efficiency measure based on robust and closest targets in DEA. Section 6 illustrates the applicability of the proposed model by using a real-life data set. Finally, some conclusions and further research directions are provided in Section 7.

## 2. Literature review

The production frontier includes numerous potential projection points, and selecting the appropriate one has been widely debated among scholars. In early DEA models, targets were determined through radial projection, and DMUs could not freely choose a benchmark for efficiency calculation (Banker et al., 1984; Charnes et al., 1978). Subsequently, scholars developed non-radial DEA models, allowing DMU to select its target based on specific criteria (Chen et al., 2015; Ruggiero & Bretschneider, 1998; Tone, 2002). Different criteria have thus given rise to various types of DEA models. These criteria include cost minimization, profit maximization, maximum distance, minimum distance, and others (Färe et al., 2006; Fare et al., 2019; Pan et al., 2024; Petersen, 2018; Petridis et al., 2017; Tone et al., 2020). Olesen and Petersen (1996) argued that the limited sample size and narrow variable variation hinder estimating a frontier with clear substitution rates, as data are typically collected for administrative purposes. Additionally, the DMU's factor structure often differs from that needed for Pareto optimality, requiring input-output substitution. Therefore, they proposed the full-dimensional efficient facet (FDEF) and the extended facet production possibility set (EFPPS) which have well-defined rates of substitution. The selection of projection targets is expressed within the framework of the primal form of the DEA model. Equivalently, in the dual form of the DEA framework, it is represented as the selection of weights or virtual prices. Just like the selection of targets, the choice of weights is guided by a variety of criteria (Aigner & Asmild, 2023; Bougnol et al., 2012; Cooper et al., 2007; Hasannasab et al., 2024; Portela & Thanassoulis, 2006).

DEA studies have emphasized the importance of closest targets, as they offer efficiency projections that can be achieved with less effort compared to other alternatives (Aparicio, 2016). Based on the methods for obtaining closest targets, research can be divided into two categories: the first involves proposing a multi-step algorithm to identify the closest target (Amirteimoori & Kordrostami, 2010; González & Álvarez, 2001), while the second achieves it directly through a Mixed Integer Linear Program (MILP) (Aparicio et al., 2007). The property of strong monotonicity introduced by Baek and Lee (2009) and early least-distance measures failed to satisfy has attracted attention from several scholars. In response, some scholars introduced the free disposable hull approach as a solution; however, the outcomes proved suboptimal (Aparicio & Pastor, 2014; Fukuyama et al., 2014). Aparicio and Pastor (2013) proposed solutions based on EFPPS to address the lacking monotonicity in least-distance DEA. Zhu et al. (2022) supported this method and proposed a MILP by simplifying the algorithm in Aparicio and Pastor (2013) to determine closest targets. However, these measures based on closest targets do not consider uncertainty from the external environment.

The integration of risk factors into DEA models has emerged as a prominent focus of recent research. These studies can be broadly categorized into two main types. The first category of studies adopts a statistical perspective, considering that the data required by DEA may exhibit randomness or incompleteness. They have proposed various DEA models to address this uncertainty, including chance-constrained DEA (Chen & Zhu, 2019), imprecise DEA (Cooper et al., 2001), fuzzy DEA (Lertworasirkul et al., 2003), bootstrapping DEA (Simar & Wilson, 1998), robust DEA (Arabmaldar et al., 2024; Bertsimas & Sim, 2004) and others (Chen et al., 2019; Ehrgott et al., 2018). The second category of studies treats data as deterministic or ignores its statistical randomness and incompleteness. In this context, uncertainty is not inherent to the data but is instead the focus of the study, often measured through methods like setting deterministic risk parameters or uncertainty ranges. Risk is reflected in prices, which has led to significant research on the uncertainty of these prices. (Camanho & Dyson, 2005; Schaffnit et al., 1997; Thompson et al., 1996; Topcu & Triantis, 2022).

Despite significant contributions from prior research, some gaps remain. First, to our knowledge, no DEA model incorporating risk factors has achieved the closest targets with well-defined substitution rates. Second, most DEA studies focus on data certainty from a statistical perspective. Although some DEA-based models incorporate risk to measure efficiency, they often fail to offer solutions for mitigating these risks from the perspective of choosing projection targets. For example, for a DMU with no prior knowledge of risk, what projection targets should be chosen to maximize risk resistance? They cannot answer the question.

### 3. Full dimensional efficient facet

Consider $N$ DMUs that all use inputs $x \in \mathbb{R}_+^M$ to produce outputs $y \in \mathbb{R}_+^S$. We define the input and output vectors of the $DMU_j$ $(j = 1, \dots, N)$ as $x_j = (x_{1,j}, x_{2,j}, \cdots, x_{m,j})^T$ and $y_j = (y_{1,j}, y_{2,j}, \cdots, y_{s,j})^T$, respectively. The production technology is denoted by $T$:

$$T \equiv \{(x, y) | y \text{ can be produced from } x\}.$$

We construct $T$ from the $N$ observations by assuming several postulates (see (Charnes et al., 1978)):

$$T \equiv \left\{ (x, y) \in \mathbb{R}_+^s \times \mathbb{R}_+^m \mid \sum_{j=1}^n \lambda_j x_j \leq x, \sum_{j=1}^n \lambda_j y_j \geq y, \lambda_j \geq 0 \right\}. \tag{1}$$

The intensity variables $\lambda_j$ is restricted to being nonnegative, indicating that the constant return to scale is imposed on the production technology.

Let $\mathcal{P}$ be defined as the polyhedral cone formed by output-input multipliers or virtual prices $(u, -v)$, derived from the half-space constraints specified in the multiplier formulation of the CCR model. Let $\mathbb{E}$ denote the set of extreme efficient DMUs. It is evident that $\mathcal{P}$ serves as the dual representation of $T$:

$$\mathcal{P} \equiv \{(u, -v) \in \mathbb{R}_+^s \times \mathbb{R}_-^m \mid u^T y_j - v^T x_j \leq 0, j \in \mathbb{E}, (u, v) \neq (0,0)\}. \tag{2}$$

The strongly efficient frontier $EffT$ is:

$$EffT \equiv \{(y, x) \in T \mid \exists (u, -v) \in \mathcal{P}, u \geq \epsilon \mathbf{1}_s, v \geq \epsilon \mathbf{1}_m : u^T y - v^T x = 0\}, \tag{3}$$

where $\mathbf{1}_s$ and $\mathbf{1}_m$ is the vectors $(1,\ldots 1)$ of dimension $s$ and $m$, and $\epsilon$ is a non-Archimedian. Formally, it is necessary to introduce some important notation. We employ a regularity condition in this study:

**Regularity Condition.** Every subset of $s+m$ columns for $DMU_j, j \in \mathbb{E}$ of the data matrix $[y_j^T, x_j^T]^T, j \in \{1,\ldots,n\}$ is affinely independent.

Then the core notation is defined:

**Definition 1.** A face qualifies as a FDEF if it is an efficient facet of dimension $d = s + m - 1$, where $s$ and $m$ represents the number of outputs and inputs, respectively.

Olesen and Petersen (1996) and Olesen and Petersen (2024) have a detailed explanation of this concept. Based on the definition of the FDEF and the regularity condition, the set of points defining the FDEF is affinely independent and non-degenerate.

Let $\mathcal{J}$ be the family of all subsets of $\mathbb{E}$, and $\mathcal{J}_{FDEF} \equiv \{J_1,\ldots,J_F\} \subseteq \mathcal{J}$ be the subset of subsets such that for $k = 1,\ldots,F$ we have the DMUs in the index set $J_k$ thus span the $k^{th}$ FDEF. $|J_k| = (s + m - 1)$ which indicates the matrix $[x_j^T, y_j^T]^T, j \in J_k$ has $s + m - 1$ linear independent vectors. A FDEF has a unique supporting hyperplane in which the FDEF is embedded. $(u_k, -v_k)$ is the unique scaled normal vector (scaled to unit length) to the $k^{th}$ FDEF and its elements are strictly positive. For any $DMU_j$, $j \in J_k$, we have $u_k^T y_j - v_k^T x_j = 0$. $(u_k, -v_k)$ are the multipliers or virtual prices of the outputs and inputs, according to which the marginal rates of substitution between inputs and/or outputs for any DMU on the $k^{th}$ FDEF can be obtained. Notably, the marginal rates of substitution of the FDEF is unique (well-defined) due to the uniqueness of scaled normal vectors, which aligns with the marginal substitution theory in economics. In other words, each FDEF corresponds to a $(u_k, -v_k)$ and a MSR different from the other FDEFs.

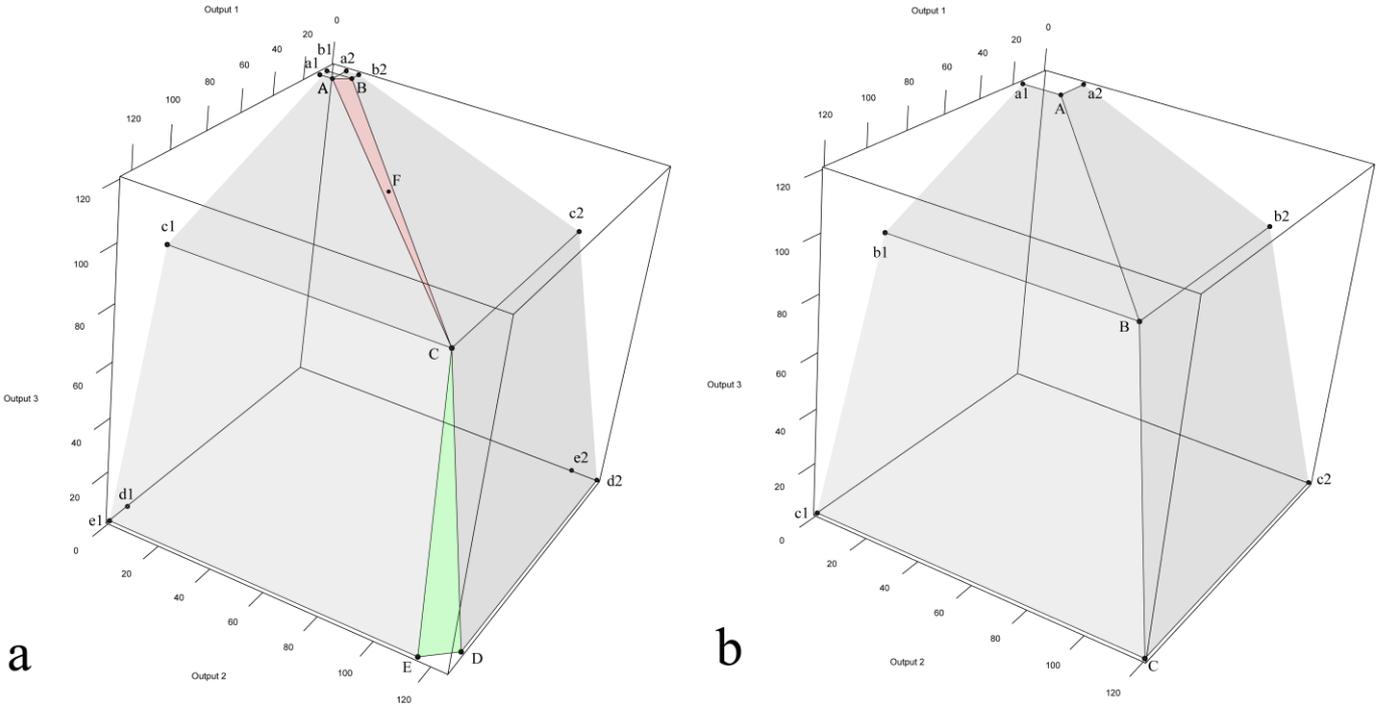

Fig. 1. The frontier including FDEFs and the frontiers without FDEFs

Table 1 Input-output data for the output isoquants in Fig. 1

|   | DMU | Input | Output1 | Output2 | Output3 |
|---|-----|-------|---------|---------|---------|
|   | A   | 1     | 5       | 10      | 120     |
|   | B   | 1     | 10      | 5       | 120     |
| a | C   | 1     | 100     | 100     | 90      |
|   | D   | 1     | 115     | 125     | 1       |
|   | E   | 1     | 125     | 115     | 1       |
|   | F   | 1     | 63      | 63      | 102     |
|   | A   | 1     | 15      | 15      | 120     |
| b | B   | 1     | 90      | 90      | 90      |
|   | C   | 1     | 120     | 120     | 1       |

We provide two examples with geometric shapes to illustrate this. For details, please refer to Fig. 1 and Table 1. Fig. 1 (a) illustrates an output isoquant that includes DMUs: A, B, C, D, E and F. These DMUs produce three outputs while utilizing an identical quantity of a single input. The technology adheres to the principle of constant returns to scale. In the given output space, the two triangles, bounded by line segments AB, BC, and CA, and by CD, DE, and EC respectively, each defines an FDEF. The vectors normal to them (scaled to unit length) are unique and strictly positive and define the marginal rates of substitution between the outputs at any point on the two FDEFs. The points labeled with lowercase letters in Fig. 1 represent the projections of A, B, C, D, and E onto other planes. Points in the output possibility set that lie beyond the two triangles do not belong to an FDEF. Consider the situation in Fig. 1 (b) with DMUs A, B and C that the two-line segments AB and BC define the efficient frontier. The marginal rates of substitution along the efficient frontier are consequently not well-defined as each segment possesses infinitely many associated price systems, including those specific to the (weakly efficient) facets they help join. The insufficient variation in the data results in the absence of a unique set of substitution rates between the outputs, and outputs 1 and 2 can only be substituted efficiently for output 3 in fixed proportions along the efficient frontier.

## 4. Single-FDEF strategy

As established above, each FDEF is associated with a unique MSR. It is important to note that these MSRs are derived by computing virtual prices or relative prices grounded in considerations of technical efficiency. The resulting virtual prices are not actual market prices; rather, they gauge the scarcity of each factor within the production process. Next, I will illustrate points located on a single FDEF exhibiting pronounced vulnerability when it comes to withstanding external risks.

For the sake of clarity in presentation, we conduct our analysis within an output-oriented DEA context where the input levels remain fixed. Hence, we have the following definitions:

Under fixed input vector $\bar{x} \in \mathbb{R}_+^m$, if the DMUs in the index set $J_k$ span the $k^{th}$ FDEF, $FDEF_k$, then

$$FDEF_k = \left\{ y \middle| y = \sum_{j \in J_k} \lambda_j y_j, \bar{x} = \sum_{j \in J_k} \lambda_j x_j, \lambda_j \geq 0 \right\} = \{y | u_k^T y - v_k^T \bar{x} = 0 \text{ and } (\bar{x}, y) \in T\}. \tag{4}$$

**Definition 2.** We define the FDEF configuration (FDEFC):

$$FDEFC = \cup_{k=1}^F FDEF_k. \tag{5}$$

Obviously, FDEFC is a set included in output possibility set with the fixed input vector $\bar{x} \in \mathbb{R}_+^m$. Then, we define the price function and revenue function. Note we only consider the impact of risk shocks on outputs under fixed inputs.

**Definition 3**. Let $\delta$ denote the risk parameter. The price vector $P(\delta) > 0$ varies with the risk parameter $\delta$, defined as:

$$P(\delta) = \big(p_1(\delta), p_2(\delta), \ldots, p_s(\delta)\big)^T \in \mathbb{R}_+^s, \tag{6}$$

where $p_i(\delta)$ represents the market price of the $i^{th}$ output. For clarity in explaining the issue, we define $\delta \in \{0,1\}$ in this section as a binary variable, where $\delta = 0$ represents a no-risk state and $\delta = 1$ indicates a fully realized risk shock.

**Definition 4.** The revenue function[1] is defined as:

---

[1] In this study, "revenue" and "market price" do not refer to narrow commercial financial gains or monetary prices but rather to broader measures of output value and evaluation weights.

$$R(y,\delta) = P(\delta)^T y = \sum_{i=1}^{s} p_i(\delta) y_i. \tag{7}$$

The term "risk" signifies that the occurrence of uncertain factors will lead to a loss in revenues. Otherwise, it would not be considered a risk but rather an opportunity. Hence, we adhere to the following assumption:

**Assumption 1.** The revenue $R(y,\delta)$ for any point exhibits a monotonically decreasing relationship with the risk parameter $\delta$. Specifically, the revenue function $R(y,\delta)$ is non-increasing with respect to $\delta$. Formally, for all feasible outputs $y \in FDEFC$ and for all $\delta \in \{0,1\}$:

$$R(y,1) \leq R(y,0). \tag{8}$$

Here, we do not assume that prices necessarily decrease due to the occurrence of risk, as some products are insensitive to risk and maintain stable prices, such as products with stable supply-demand relationships, like salt. Some products may even experience price increases when other products react to risk with price decreases. However, it is necessary to impose constraints on revenue. Because all prices are strictly positive, if $y_1 \geq y_2$ component-wise (with at least one strict inequality, then $R(y_1,\delta) > R(y_2,\delta)$, ensuring strict monotonicity of the revenue function.

**Definition 5**. The optimal revenue point $y_k^*(\delta)$ within a single FDEF, $FDEF_k$, is defined as:

$$y_k^*(\delta) = arg \max_{y \in FDEF_k} R(y,\delta), \tag{9}$$

**Definition 6.** The globally optimal revenue point $y^*(\delta)$ within the $FDEFC$ is defined as:

$$y^*(\delta) = arg \max_{y \in FDEFC} R(y,\delta), \tag{10}$$

$y_k^*(0)$ is the optimal output when there are no shocks in the market.

Regarding the number of optimal revenue points, we have the following theorem:

**Theorem 1**. Under the single $FDEF_k$, the optimal revenue output $y_k^*(\delta)$ for each $\delta \in \{0,1\}$ is not unique. Formally, for each $\delta \in \{0,1\}$,

$$y_k^*(\delta) = arg \max_{y \in FDEF_k} R(y,\delta)$$

not unique.

**Proof.** There are at least two cases where $y_k^*(\delta)$ has multiple values.

Case 1: $P(\delta) \parallel (u_k, -v_k)$.

Let $(u_k, -v_k)$ be the normal vector of $FDEF_k$ i.e., $FDEF_k$ lies in the hyperplane $\{y \in T | u_k^T y = v_k^T \bar{x} = \alpha\}$ for the constant cost $\alpha$ because we just focus on the input-oriented output possibility set.

If $P(\delta) = \lambda(u_k, -v_k)$ with $\lambda \neq 0$, then for every $y \in FDEF_k$:

$$R(y,\delta) = P(\delta)^T y = \lambda(u_k, -v_k)^T y = \lambda\alpha,$$

a constant. Hence, $y \in FDEF_k$ yield the same revenue, so the optimal solution set is the entire facet, implying $y_k^*(\delta)$ is not unique at all.

Case 2: $P(\delta) \cdot (B - A) = 0$ for a 1D sub-edge $\overline{AB} \subset FDEF_k$.

Suppose $\overline{AB}$ is a one-dimensional face (edge) within $FDEF_k$. Parametrize points in $\overline{AB}$ by

$$y(t) = A + t(B - A), \quad t \in [0,1].$$

Then, if $P(\delta) \cdot (B - A) = 0$,

$$R(y(t), \delta) = P(\delta) \cdot [A + t(B - A)] = P(\delta) \cdot A + t[P(\delta) \cdot (B - A)] = P(\delta) \cdot A.$$

Hence, every point on the entire segment $\overline{AB}$ has the same revenue $P(\delta) \cdot A$. If that value is the maximum within $FDEF_k$, then the entire line segment is optimal, so $y_k^*(\delta)$ is not unique. In both cases, we obtain multiple solutions that tie for the maximum. Thus, if $P(\delta) \cdot (B - A) = 0$ occurs, the output $y_k^*(\delta)$ achieving $\max_{y \in FDEF_k} R(y, \delta)$ cannot be uniquely determined. The theorem is proved.

**Theorem 2.** There may exist multiple FDEFs (whether adjacent or not) meeting that for each $FDEF_k, k = 1, \ldots n$, there exists a point $y_k \in FDEF_k$ satisfying

$$R(y_k, \delta) = \max_{y \in FDEFC} \{P(\delta)^T \cdot y\}.$$

In other words, although $FDEF_1, FDEF_2, \ldots, FDEF_n$ are non-parallel, there may exist a price vector by a risk parameter $\delta$ under which each FDEF possesses a distinct point that attains the same maximum revenue.

**Proof.** Based on the proposition of the non-uniqueness of the optimal revenue point within a single FDEF, we can conclude that points on a common edge shared by multiple FDEFs may all be potential points of maximum revenue, thus aligning with this proposition. However, this holds under the condition that these FDEFs are adjacent. Our focus is to analyze whether this proposition still holds true even when the FDEFs are not adjacent.

Based on the defined properties of FDEFs, any two distinct FDEFs, $FDEF_j$ and $FDEF_k (j \neq k)$, have normal vectors $(u_j, -v_j)^T$ and $(u_k, -v_k)^T$ that are not proportional, i.e, there does not exist a scalar $\lambda > 0$ such that $(u_j, -v_j)^T = \lambda (u_k, -v_k)^T$. This directly implies that any two distinct FDEFs are non-parallel geometrically. Thus, the FDEFC consists of multiple non-parallel FDEFs and is not necessarily convex. Hence, the price vector $P(\delta)$ can be designed or exists such that it supports multiple FDEFs at different points simultaneously, each achieving the same maximum yield. In other words, the revenue function can have a hyperplane that touches multiple FDEFs. The theorem is proved.

Olesen and Petersen (1996) introduces an FDEFC, as shown in Table 2 and Figure 2B of that paper. This FDEFC is composed of two FDEFs (one formed by DMU1, 2, and 3, and the other by DMU5, 6, and 7). Clearly, by the line connecting DMU3 and DMU5, which is tangent to both FDEFs, there exists a hyperplane passing through this line.

Our analysis is conducted in a two-stage context, distinguishing between the periods before and after the onset of risk.

Prior to the emergence of risk, the inefficient observed DMU, $DMU_o$, sets a target $\hat{y} \in FDEF_k$ and successfully attains it. $\hat{y}$ have the revenue: $R(\hat{y}, 0)$. After the risk emerges, the DMU adjusts its outputs starting from $\hat{y}$, utilizing the MSR corresponding to the $FDEF_k$, to ultimately reach the optimal revenue point, $y_k^*(1)$, of which the revenue is $R(y_k^*(1), 1)$. $\hat{y}$ is $y_k^*(0)$ if the inefficient $DMU_o$ aim to maximize the revenue. However, in real production, the choice of a target is influenced by various considerations, including minimizing the distance to the benchmark (Aparicio & Pastor, 2014; Zhu et al., 2022), maximizing the slack (Tone, 2002), minimizing cost (Ang et al., 2019) or adhering to predefined directional vectors (Chung et al., 1997). Hence, we do not use $y_k^*(0)$ instead of $\hat{y}$ directly. Our research focuses on countering risk, so we must assume that maximizing revenue after a risk shock is the primary objective considered by $DMU_o$.

Risks always cause DMUs to incur losses, even if measures can be taken to reduce those losses. As the saying goes, "You cannot stop the waves, but you can learn to surf" highlighting that while risks are unavoidable, efforts can be made to minimize their impact. Thus, we follow the following assumption:

**Assumption 2.** While the loss can be partially or even fully mitigated through output allocation, the post-risk revenue, $R(y_k^*(1), 1)$, do not exceed the original revenue, $R(\hat{y}, 0)$. Formally,

$$R(y_k^*(1), 1) \leq R(\hat{y}, 0). \tag{11}$$

This inequality states that the adjusted revenue under the risk scenario ($\delta = 1$) cannot surpass the original revenue under the no-risk scenario ($\delta = 0$). If $R(y_k^*(1), 1) = R(\hat{y}, 0)$, the loss is fully mitigated; otherwise, if $R(y_k^*(1), 1) < R(\hat{y}, 0)$, only partial mitigation is achieved. The loss is $R(\hat{y}, 0) - R(y_k^*(1), 1)$. In no event can the post-risk revenue surpass the pre-risk revenue. If the $DMU_o$ does not engage in output substitution, it will incur a loss of $R(\hat{y}, 0) - R(\hat{y}, 1)$. Therefore, by leveraging the inherent MSR, a loss recovery of $R(y_k^*(1), 1) - R(\hat{y}, 1)$ can be achieved.

**Definition 6**. The single FDEF's capacity to withstand risk is defined as:

$$WR = R(y_k^*(1), 1) - R(\hat{y}, 1). \tag{12}$$

$R(y_k^*(1), 1) - R(\hat{y}, 1) = 0$ indicates that the initial projection point is already the optimal revenue point after the risk occurs ($y_k^*(1) = \hat{y}$), meaning the $DMU_o$ cannot reduce the loss caused by the risk through substitution relationship.

Based on the above assumption, we can derive the following important theorems:

**Theorem 3**. Under the assumption 2, the globally optimal revenue point $y^*(1) = \arg\max_{y \in FDEFC} R(y, 1)$ cannot achieve revenue that exceeds its pre-risk level, $R(\hat{y}, 0)$. Formally:

$$R(y^*(1), 1) \leq R(\hat{y}, 0). \tag{13}$$

**Proof.** Suppose, for contradiction, that the globally optimal revenue point under risk, $y^*(1)$, achieves a revenue greater than the pre-risk revenue:

$$R(y^*(1), 1) > R(\hat{y}, 0).$$

Since $y^*(1) \in FDEFC$, it must lie in at least one of the FDEFs, so we suppose:

$$y^*(1) \in FDEF_{k^*}.$$

Inside this single $FDEF_{k^*}$, the optimal revenue point must exist:

$$y_{k^*}^*(1) = arg \max_{y \in FDEF_{k^*}} R(y, 1) = y^*(1).$$

Then, we must have:

$$R(y^*(1), 1) = R(y_{k^*}^*(1), 1).$$

From the assumption 2 we know:

$$R(y_{k^*}^*(1), 1) \leq R(\hat{y}, 0).$$

This implies:

$$R(y^*(1), 1) \leq R(\hat{y}, 0).$$

This contradicts our initial supposition that $R(y^*(1), 1) > R(\hat{y}, 0)$. Hence, we have $R(y^*(1), 1) \leq R(\hat{y}, 0)$. The theorem is proved.

The assumption 2 limits the revenue brought about by output adjustment or the capacity to withstand risk of a single FDEF. Thus, we have:

**Theorem 4.** Then there exists an upper bound for $WR$:

$$WR \leq R(\hat{y}, 0) - R(\hat{y}, 1). \tag{14}$$

**Proof.**

$$WR = R(y_k^*(1), 1) - R(\hat{y}, 1).$$

Since from the assumption 2, we have:

$$R(y_k^*(1), 1) \leq R(\hat{y}, 0).$$

Substitute this upper bound into the expression for $WR$:

$$WR = R(y_k^*(1), 1) - R(\hat{y}, 1) \leq R(\hat{y}, 0) - R(\hat{y}, 1).$$

Thus, we have obtained:

$$WR \leq R(\hat{y}, 0) - R(\hat{y}, 1).$$

The theorem is proved.

The difference $R(\hat{y}, 0) - R(\hat{y}, 1)$ can be viewed as the initial "risk-induced revenue gap" at $\hat{y}$. Since $y_k^*(1)$ is chosen to maximize the revenue within the single FDEF, $WR$ represents how much better the $DMU_o$ does compared to staying at $\hat{y}$ under risk. The fact that $WR \leq R(\hat{y},0) - R(\hat{y},1)$ means that the DMU cannot recover more than the entire lost revenue at $\hat{y}$. This sets a strict upper limit on how much improvement is achievable by the MSR of a single FDEF.

The next theorem is one of the core elements of this study, which reflects a single FDEF's pronounced vulnerability when it comes to withstanding risks. Additionally, the robust projection target proposed in the following sections is based on this theorem.

**Theorem 5.** For any $\delta \in \{0,1\}$, the following inequality holds:

$$R(y_k^*(\delta), \delta) \leq R(y^*(\delta), \delta). \tag{15}$$

This implies that the optimal revenue within a single FDEF does not exceed the globally optimal revenue.

**Proof.** By definition, $y_k^*(\delta)$ is the optimal point within the single $FDEF_k$, satisfying:

$$y_k^*(\delta) = arg \max_{y \in FDEF_k} R(y, \delta).$$

On the other hand, $y^*(\delta)$ is the globally optimal revenue point defined as:

$$y^*(\delta) = arg \max_{y \in FDEFC} R(y, \delta).$$

Since $FDEF_k \subseteq FDEFC$, it follows that:

$$R(y_k^*(\delta), \delta) \leq \max_{y \in FDEFC} R(y, \delta) = R(y^*(\delta), \delta)$$

Thus, the theorem is proved.

For the $DMU_o$ on the single $FDEF_k$, the risk loss that cannot be mitigated through the MSR is $R(y^*(1),1) - R(y_k^*(1),1) \geq 0$. If $R(y^*(1),1) - R(y_k^*(1), 1) = 0$, that indicates $y^*(\delta) = y_k^*(\delta)$, which is the ideal situation. However, this is typically not easy to achieve. The optimal point on the $FDEF_k$ is not necessarily the globally optimal point, indicating the limitation of a single FDEF in mitigating the impact of risk shocks. While the allocation of resources within a single FDEF can reduce the negative impact of risk on revenue, there may still be some losses that can be further reduced by reaching the globally optimal revenue point in the FDEFC. In other words, after the risk has materialized, the globally optimal revenue point within the FDEFC may not lie on the FDEF where the DMU currently resides. This highlights the need to diversify alternative MSRs to enhance adaptability in a risky environment.

Next, we will use an example to illustrate the limit of a single FDEF when confronted with risk. We use the DMUs in Table 1 as samples and know that there is a total of two FDEFs in Fig.1. We let the subset $J_1 = \{A, B, C\}$ in which the DMUs span the $FDEF_1$ and the subset $J_2 = \{C, D, E\}$ in which the DMUs span the $FDEF_2$. Let the market price function vectors of the output 1, output 2, and output 3 are:

$$P(\delta) = \big(p_1(\delta), p_2(\delta), p_3(\delta)\big)^T = (5, 5 + 5\delta, 12 - 11.9\delta)^T.$$

It can be observed that the price of output 1 is unaffected by external risks, similar to products with stable supply-demand relationships, such as salt. The price of output 2 increases with market risk shocks, such as heightened demand for umbrellas due to rainy weather. In contrast, output 3 is highly negatively sensitive to market shocks, such as the restaurant industry during an avian flu outbreak. These three different market price functions align perfectly with real-world scenarios. The price function here aligns with the assumption previously stated that prices necessarily decrease due to the occurrence of risk. We just set that the revenue function is non-increasing with respect to $\delta$.

We assume that, in the initial state, the market is tranquil, and therefore $\delta = 0$, and the market prices are: $P^0 = (p_1^0, p_2^0, p_3^0) = (5, 5, 12)$. We consider point F as the initial target or production state. The revenue at point F, $R(F, 0)$, equals to 1854. When the market is subjected to shocks ($\delta = 1$), the market prices are: $P^1 = (p_1^1, p_2^1, p_3^1) = (5, 10, 0.1)$. The revenue at point F, $R(F, 1)$, equals to 955.2 now. Hence, the corresponding loss in revenue is equal to $1854 - 955.2 = 898.8$. As mentioned above, points on the FDEF can resist some degree of risk based on the inherent substitution relationship. Therefore, the $DMU_o$ can leverage the internal substitution relationship to shift from point F to other points, adapting to the scenario where output 3 is undervalued by the market.

We find that D is the globally optimal revenue point within the FDEFC formed by $FDEF_1$ and $FDEF_2$ when $\delta = 1$. It can be observed that D is the point with the highest allocation efficiency under risk shocks across the two FDEFs, and $R(D, 1) = 1825.1$. If $DMU_o$ could move directly from F to D, the loss is $(1854 - 1825.1) = 28.9$, which would almost eliminate the negative impact of the risk. However, the $DMU_o$ can only move to the optimal revenue point on the $FDEF_1$ under the prices $(5, 10, 0.1)$, point C, based on the substitution relationship determined by $FDEF_1$. Note the two FDEFs correspond to two distinct MSRs, essentially representing two different production and operational models. Although the DMU can reach point C through the MSR of $FDEF_1$, and C also belongs to $FDEF_2$, this does not automatically imply that it adopts the business operation logic of $FDEF_2$. Since the DMU is already operating within the production model of $FDEF_1$, transitioning between two mature production models in practice may disrupt normal operations. Therefore, this study assumes that such a transition is not feasible.

Under certain conditions, the optimal revenue point may not be unique and could be any point on the $FDEF_1$. The revenue $R(C, 1) = 1509$, which is an 58% increase compared to $R(F, 1)$. The corresponding loss is $(1854 - 1509) = 345$. By relying solely on the MSR within the FDEF, it is possible to reduce the loss by 553.8. However, there is still a loss of 345, which, in this case, cannot be eliminated within the $FDEF_1$ framework. It is an unavoidable loss that must be borne. However, there is still significant potential for loss reduction (moving to point D) that remains untapped, which demonstrates the limitations of the single-FDEF strategy. It is important to note that the MSR within the FDEF does not always significantly mitigate risk. The extent of mitigation depends on the DMU's initial position and specific price functions.

## 5. Multi-FDEF intersection strategy

When a DMU selects a benchmark, it is effectively choosing the normal vector or MSR associated with the FDEF where the benchmark lies. The DMU does not know in advance when risk will occur or what the market price functions will be under risk shocks. Thus, it cannot be determined whether the optimal revenue point reached through the MSR within the single FDEF aligns with the globally optimal revenue point within the FDEFC after the risk materializes. This outcome can only be determined once the risk occurs. From a risk mitigation perspective, the DMU should choose a MSR that maximizes the likelihood of reaching the globally optimal revenue point within the FDEFC under post-risk conditions.

The intersection points of multiple FDEFs meet the criteria for an ideal projection target. Since an intersection point belongs to each of the FDEFs involved, it can select the corresponding MSR (virtual prices) from any of these FDEFs, enabling the DMU to potentially reach any position within any of these facets. After a risk shock, based on the prevailing market prices at that time, the DMU can choose the point located on these FDEFs that yields the greatest revenues. Hence, we have the following theorem:

**Theorem 6.** A point at a multi-FDEF intersection has a greater probability of moving to the globally optimal point under any realization of $\delta$ than one within a single FDEF (not on the intersection of the single FDEF with other FDEFs).

**Proof.** $\delta$ is a random variable representing the state of risk. Note at this point, $\delta$ is no longer a binary variable like in the last section but instead takes multiple values. $P(\delta)$ is the price vector function corresponding to $\delta$. The globally optimal point under the risk scenario $\delta$ is:

$$y^*(\delta) = arg \max_{y \in FDEFC} R(y, \delta).$$

Define events:

$$A_k = \{\delta \in D: y_k^*(\delta) \in FDEF_k\}, \quad k = 1, \ldots, F.$$

$\mathbb{P}(A_k)$ is the probability of achieving the globally optimal point in a single FDEF (say the $k$-th one). $\{A_k\}$ are not pairwise disjoint, there may exist $\delta_0$ such that

$$\delta_0 \in A_i \cap A_j, \quad i \neq j,$$

meaning

$$y^*(\delta_0) \in FDEF_i \cap FDEF_j.$$

The global optimum, at $\delta_0$, simultaneously resides on multiple FDEFs (common intersection). Extending this logic, the same $\delta_0$ might lie in the intersection of 3 or more FDEFs as well.

If the DMU, before the shock is realized, chooses only the MSR corresponding to the single FDEF, then, the probability of achieving the globally optimal point is $\mathbb{P}(A_k)$ after the shock is realized. This represents the DMU's probability of successfully matching the chosen FDEF and the global optimum.

Now, if there exists a point $y^\dagger$ such that:

$$y^\dagger \in \bigcap_{k=1}^{n} FDEF_k, n > 1,$$

then the DMU, before $\delta$ is realized, effectively secures $n$ FDEF-based MSRs for future use. Define the event:

$$B = \bigcup_{k=1}^{n} A_k.$$

Then,

$$A = A_k \subseteq \bigcup_{k=1}^{n} A_k = B, k = 1, \ldots, n$$

Under non-disjointness, $\mathbb{P}(B)$ must be computed using the Vermph (inclusion-exclusion principles).

$$\mathbb{P}(B) = \sum_{k=1}^{n} \mathbb{P}(A_k) - \sum_{1 \leq i < j \leq n} \mathbb{P}(A_i \cap A_j) + \cdots.$$

If multiple FDEFs are simultaneously optimal for some $\delta$, then certain intersection terms are strictly positive. Comparing the single-FDEF strategy's $\mathbb{P}(A_k)$ with the multi-FDEF intersection strategy's $\mathbb{P}(B)$, even if events overlap, we have:

$$\mathbb{P}(B) \geq \max_{1 \leq k \leq n} \mathbb{P}(A_k).$$

Therefore, when the DMU obtains $n > 1$ FDEF options via a multi-FDEF intersection point $y^\dagger$, the probability of post-risk alignment with the global optimum increases to $\mathbb{P}(B)$, which is certainly not inferior to any single $\mathbb{P}(A_k)$, and generally strictly greater. The theorem is proved.

**Theorem 7**. Let one intersection point be $y_1^\dagger \in \bigcap_{k \in K_1} FDEF_k$, with index set $K_1 \subseteq \{1, \ldots, F\}$ and $|K_1| = n_1$. Another intersection point $y_2^\dagger \in \bigcap_{k \in K_2} FDEF_k$, with index set $K_2 \subseteq \{1, \ldots, F\}$ and $|K_2| = n_2$. If $K_1 \subseteq K_2$ (ie, $|K_2| \geq n_1$ and the facets in $K_2$ include those in $K_1$), then adopting the intersection point $y_2^\dagger$ is at least as robust, and generally more robust in terms of risk success probability, than the intersection point $y_1^\dagger$ involving fewer FDEFs ($K_1$). Therefore, the most robust multi-FDEF intersection point is $y^\dagger \in \bigcap_{k \in K^*} FDEF_k$, where $|K^*|$ is maximal over all subsets $K \subseteq \{1, \ldots, F\}$ with $\bigcap_{k \in K} FDEF_k \neq \emptyset$.

**Proof.** For $K_1$, define $B_1 = \bigcup_{k \in K_1} A_k$. If the DMU can switch to any facet in $K_1$, it succeeds whenever $\delta \in B_1$. For $K_2$, define $B_2 = \bigcup_{k \in K_2} A_k$. If $K_1 \subseteq K_2$, then $B_1 = \bigcup_{k \in K_1} A_k \subseteq \bigcup_{k \in K_2} A_k = B_2$, implying $\mathbb{P}(B_1) \leq \mathbb{P}(B_2)$. Since the sets $\{A_k\}$ can overlap, $\mathbb{P}(B_1)$ and $\mathbb{P}(B_2)$ must be computed via inclusion-exclusion: $\mathbb{P}\left(\bigcup_{k\in K_i} A_k\right) = \sum_{k\in K_i}\mathbb{P}(A_k) - \sum_{k_1<k_2\in K_i}\mathbb{P}(A_{k_1}\cap A_{k_2}) + \dots$ Hence, the containment $B_1 \subseteq B_2$ remains valid regardless of overlaps. Consequently, among various multi-FDEF intersection points, the one involving the greatest number of facets provides the highest (or equal) success probability under risk, thus the highest robustness. The most robust multi-FDEF intersection point is $y^\dagger \in \bigcap_{k\in K^*} FDEF_k$, where $|K^*|$ is maximal over all subsets $K \subseteq \{1, \dots, F\}$ with $\bigcap_{k\in K} FDEF_k \neq \emptyset$. The theorem is proved.

Without $K_1 \subseteq K_2$ (or vice versa), the fact that $|K_2| \geq |K_1|$ alone does not guarantee that the intersection point belonging to more FDEFs is more robust. Event coverage and probability outcomes depend on the underlying distribution $\mathbb{P}(\cdot)$ and how the sets $\{A_k\}$ intersect. There is a possibility that some of the FDEFs in $K_2$ could overlap or have correlations that do not actually expand coverage enough to exceed $\mathbb{P}(B_1)$. This research focuses on examining the optimal projection targets of a DMU under conditions where no prior information regarding risks is available. Hence, we do not assume a probability distribution for post-risk prices. If the DMU has no risk information, the optimal pre-risk decision should be the multi-FDEF intersection strategy rather than any single-FDEF strategy. While this approach may still fail to achieve the globally optimal revenue, it maximizes the likelihood of success. Thus, we have the important definition of robust point:

**Definition 7.** A point is robust if it belongs to multi-FDEF intersection, $y^\dagger \in \bigcap_{k\in K^*} FDEF_k$, and $|K^*|$ is maximal over all subsets $K \subseteq \{1, \dots, F\}$ with $\bigcap_{k\in K} FDEF_k \neq \emptyset$.

It can also be understood in this way. The intersection point allows the DMU to adapt by selecting the most favorable FDEF and the associated MSR, ensuring improved performance in a risk environment. Furthermore, without prior knowledge of the specific risk, the DMU can leverage these MSRs to cover as many positions as possible along the production frontier, thereby maximizing its adaptability and strategic flexibility even in the face of uncertainty.

The example in the last section illustrates the limit of single FDEF by restricting the $DMU_o$ from moving to the globally optimal revenue point D. The multi-FDEF intersection strategy means the $DMU_o$ should select the intersection point C of the two FDEFs ($FDEF_1$ and $FDEF_2$) as the projection target. Point C, as the intersection of two FDEFs, allows the DMU positioned at C to determine the globally optimal revenue point D on the $FDEF_2$ based on post-risk prices. The DMU can then adjust its output using the corresponding MSR to reach point D. Then, the loss is $(1854 - 1825.1) = 28.9$, which would nearly eliminate the adverse effects of the risk. If the globally optimal revenue point lies on the $FDEF_1$, assumed to be A, the DMU can similarly adopt the MSR of the $FDEF_1$ to move to A. These highlight the superiority of this strategy over the single-FDEF strategy in managing risks.

## 6. DEA based on robust and closest targets

We have demonstrated the validity of multi-FDEF intersection strategy in managing risks in the last section. Additionally, the intersection of multiple FDEFs could be a line segment, or in higher-dimensional spaces, even a plane or a cube. One issue is which of the many robust points should be chosen as the projection target. The closest targets closely resemble the observed inputs and outputs of the evaluated DMU and require less effort to achieve compared to other options, such as the farthest benchmarking. Thus, we propose an efficiency measure based on robust and closest targets.

The set of feasible outputs corresponding to the strongly efficient frontier $EffO$ is:

$$EffO \equiv \{y \mid \exists (u, -v) \in \mathcal{P}, u \geq \epsilon \mathbf{1}_s, v \geq \epsilon \mathbf{1}_m : u^T y - v^T x_o = 0\}. \tag{16}$$

This measure is based on the Russell output measure. Thus, let us give the original Russell output inefficiency measure with the aim of determining the closest targets for any observed $DMU_o, (x_o, y_o)$:

$$\Gamma(x_o, y_o) = Min \left\{ \frac{1}{s} \sum_{r=1}^{s} \frac{s_r}{y_{ro}} : (y_o + s) \in EffO, s = (s_1, \dots, s_s) \geq \mathbf{0}_s \right\}. \tag{17}$$

The preceding model can be implemented in DEA as the following linear programming:

$$\begin{aligned}
& Min \frac{1}{s} \sum_{r=1}^{s} \frac{s_r}{y_{ro}} \\
& s.t. \\
& \sum_{j=1}^{n} \lambda_j x_{ij} \leq x_{io}, i = 1, \dots, m. \\
& \sum_{j=1}^{n} \lambda_j y_{rj} = y_{ro} + s_r, r = 1, \dots, s. \\
& \lambda_j \geq 0, j = 1, \dots, n. \\
& s_r \geq 0, i = 1, \dots, s.
\end{aligned} \tag{18}$$

It should be noted that when we let the $FDEF_k$ be the benchmark, one special situation may arise. For example, we choose the FDEF, the triangle ABC, as the benchmark and measure the efficiency of the point D. It is easy to find that C is the projection of D. The outputs of C and D are (100,100,90) and (115,125,1). The output1 and output2 of D are more than the two outputs of C. Hence, the slacks of the two outputs for D are less than 0. Therefore, when we let the $FDEF_k$ be the benchmark we cannot define Russell output inefficiency measure as:

$$\Gamma(x_o, y_o) = Min \left\{ \frac{1}{s} \sum_{r=1}^{s} \frac{s_r}{y_{ro}} : (y_o + s) \in FDEF_k, s = (s_1, \dots, s_s) \right\} \tag{19}$$

which is equivalent to

$$Min \frac{1}{s}\sum_{r=1}^{S} \frac{s_r}{y_{ro}}$$
$$s.t.$$
$$\sum_{j=1}^{|J_k|} \lambda_j x_{ij} \leq x_{io}, i = 1,\dots,m.$$
$$\sum_{j=1}^{|J_k|} \lambda_j y_{rj} = y_{ro} + s_r, r = 1,\dots,s$$
$$\lambda_j \geq 0, s_r \geq 0, j \in J_k.$$
(20)

That is because the programming may have no solution. Therefore, we propose the following model that effectively addresses this issue:

$$Min\ W \sum_{r=1}^{S}(1-z_r) + \frac{1}{s}\sum_{r=1}^{S}\frac{s_r^+ + s_r^-}{y_{ro}}$$
$$s.t.$$
$$\sum_{j=1}^{|J_k|} \lambda_j x_{ij} \leq x_{io}, i = 1,\dots,m$$
$$\sum_{j=1}^{|J_k|} \lambda_j y_{rj} = y_{ro} + s_r^+ - s_r^-, r = 1,\dots,s$$
$$s_r^+ - s_r^- = s_r, r = 1,\dots,s$$
$$s_r \geq -M(1-z_r), r = 1,\dots,s$$
$$s_r \leq M z_r, r = 1,\dots,s$$
$$s_r^+ \geq 0, r = 1,\dots,s$$
$$s_r^- \geq 0, r = 1,\dots,s$$
$$z_r \in \{0,1\}.$$
(21)

Then, the corresponding efficiency is:

$$\theta(x_o, y_o) = \frac{1}{1 + \frac{1}{s}\sum_{r=1}^{s}\frac{s_r^{+*} + s_r^{-*}}{y_{ro}}}$$
(22)

Note that this formula allows the slacks, $s_r^+ - s_r^- = s_r$, of outputs to be negative, to accommodate the special circumstances mentioned above. The most DEA model requires that slack variables must be non-negative to align with economic principles, but the negative slack in our model has economic significance. Negative slack signifies that the output of the observed DMU surpasses that of the robust projected target, necessitating a reduction in output for the observed DMU to reach the robust projected target. Hence, negative slack measures the degree of distortion in resource allocation leading to the shortfall in the ability to withstand risks. Positive slack measures the output that the observed DMU currently cannot achieve, like other DEA models. In the Russell model, minimizing the weighted average of the proportion of the positive slack relative to its corresponding output can identify the least-distance benchmark. In our model allowing negative slack, we take the absolute value of the slack variables, which is equal to $s_r^+ + s_r^-$ in the second term of the objective function. In other words, when defining distance, we still take negative slack into consideration.

Meanwhile, we apply the fourth and fifth constraints to the slack using the Big-M method, which ensures: when $z_r = 1$, then $M \geq s_r \geq 0$ (allowing positive values); when $z_r = 0$, then $0 \geq s_r \geq -M$ (allowing negative values). $M$ is a sufficiently large positive number, and $z_r$ is a binary variable. In the objective function, the weight $W \gg 1/s > 0$, and we set it equal to 10000 in this study. Hence, the model will prioritize minimizing $\sum_{r=1}^{s}(1 - z_r)$, which is equivalent to maximizing $\sum_{r=1}^{s} z_r$ and ensuring that the slacks $s_r$ are as much as possible positive number or zero. It is necessary to make such a setting. The parameter $\lambda_j$ can take any value greater than 0, and the slack variables are allowed to be negative. This flexibility allows the model to select any input-output combination at any scale as a benchmark, and different scales correspond to different values of the slack variables, including negative ones. While we recognize the rationale for permitting negative slack, its application is introduced solely in exceptional and specific circumstances. Therefore, we set the optimization goal of ensuring that the slack variables are as many as possible non-negative as the priority. Otherwise, driven by the objective function, the observed DMU may strategically select a small $\lambda_j$ to attain a projection target with low outputs leading to substantial negative slack.

To develop a robust efficiency measure, we emphasize that benchmark selection should be guided not only by resource efficiency but also by resilience to risk. Therefore, we can select all the robust points as the benchmark to ensure robust efficiency. While the output at the projection may be lower than that of the evaluated DMU, this can be effectively managed using the model we have just introduced. Notably, the robustness of the projection point is our primary consideration, with the minimum projection distance serving as a secondary factor.

Now we need determine all the robust points, that is $y^\dagger \in \bigcap_{k \in K^*} FDEF_k$, where $|K^*|$ is maximal over all subsets $K \subseteq \{1, \ldots, F\}$ with $\bigcap_{k \in K} FDEF_k \neq \emptyset$. Note that these points are not necessarily all adjacent; they may form several non-adjacent faces. In other words, some DMUs may participate in the same maximum number of FDEFs, but the specific FDEFs they are involved in could differ. Each FDEF is spanned by the extreme efficient DMUs we have identified. However, the robust point closest to the observed DMU is not necessarily one of these extreme efficient DMUs; it could be a point on the convex hull formed by the extreme efficient DMUs within a single FDEF. Note that extreme efficient DMUs from different FDEFs cannot form a convex hull together. Therefore, we need to group the robust points based on the specific FDEFs they collectively contribute to forming. We propose an algorithm to determine and partition all the robust points.

There are $F$ FDEFs in sample, denoted by $FDEF_k$, $k = 1, \ldots, F$. $\mathcal{D} = \{d_1, d_2, \ldots, d_N\}$ is the set of all DMUs spanning all the FDEFs. For each $FDEF_k$, a subset $J_k \subseteq \mathcal{D}$ representing the DMUs that construct it, where $|J_k| = s +$

$m - 1$.

Consider the following algorithm *Determine and partition*:

Step 1. Prepare a mapping to record, for each DMU $d$, the set of FDEF it belongs to. Build the Function $K(d)$ and let $K(d) \leftarrow \emptyset$. For each FDEF, $FDEF_k$, $k \in \{1,2,\ldots,F\}$: retrieve the set $J_k$, i.e., the DMUs that construct $FDEF_k$; For each DMU $d$ in $J_k$, update

$$K(d) \leftarrow K(d) \cup \{FDEF_k\}.$$

After this loop, $K(d)$ contains all the FDEFs in which DMU $d$ participates.

Step 2. Compute participation counts and find the maximum. For each $d \in \mathcal{D}$, define

$$Count(d) = |K(d)|.$$

Determine

$$Maxcount = \underset{d \in \mathcal{D}}{Max}\ Count(d).$$

Step 3. Identify $S^*$: the set of DMUs with maximal FDEF participation

$$S^* = \{d \in \mathcal{D}\ |\ Count(d) = Maxcount\}.$$

Step 4. Partition $S^*$ by the exact subset of FDEFs. Create a mapping:

$$GroupMap:\ (\text{subset of FDEFs}) \rightarrow \text{List of DMUs}.$$

For each $d \in S^*$: Let $G_d = K(d)$ and insert $d$ into $GroupMap\ [G_d]$. (If $GroupMap\ [G_d]$ does not exist yet, initialize it). As a result, each distinct subset of FDEFs, $G_d$, in GroupMap corresponds to one "group". The total number of groups is determined by the number of different keys in $GroupMap$ or distinct subsets of FDEFs observed within $S^*$:

$$GroupsCount = |\{G_d\ |\ d \in S^*\}|.$$

Suppose there are $H$ distinct $G_d$, denoted by $G_p, p = 1,\ldots,H$, and we have:

$$S_p^* = \{d \in S^*\ |\ K(d) = G_p\},$$

which includes all DMUs in $S^*$ whose corresponding subset of FDEFs is exactly $G_p$. $\cup_{p \in \{1,\ldots,H\}} S_p^* = S^*$ means that DMUs in $S^*$ are divided into $H$ groups. Each subset of the robust points, $S_p^*$, correspond to $T(S_p^*)$, the production possibility set formed by these extreme efficient DMUs in $S_p^*$. Formally,

$$T(S_p^*) = \left\{(x,y) \in \mathbb{R}_+^s \times \mathbb{R}_+^m\ \bigg|\ \sum_{j \in J_p} \lambda_j x_j \leq x, \sum_{j \in J_p} \lambda_j y_j \geq y, \lambda_j \geq 0\right\},$$

which can be part of the production frontier. $J_p = \{j: d_j \in S_p^*\}$, $p = 1,\ldots,H$ is the index set of DMUs included in group $S_p^*$. In addition, $\cup_{p \in \{1,\ldots,H\}} T(S_p^*)$ is the entire production frontier.

This algorithm *Determine and partition* filters the robust points for risk mitigation under the multi-FDEF intersection strategy from the set of all FDEFs. These points are then used as the frontier for minimum-distance Russell model, serving as candidates for the closest projection point.

The procedure for measuring robust efficiency based closest targets consists of four steps:

Step 1: Determine the set of extreme efficient units by solving the following model for each DMU in the sample:

$$
\begin{aligned}
& Min \lambda_0 \\
& s.t. \\
& \sum_{j=1}^{n} \lambda_j x_{ij} \leqslant x_{io}, i = 1, \ldots, m \\
& \sum_{j=1}^{n} \lambda_j y_{rj} \geqslant y_{ro}, r = 1, \ldots, s. \\
& \sum_{j=1}^{n} \lambda_j = 1, \\
& \lambda_j \geqslant 0, j = 1, \ldots, n.
\end{aligned} \quad (23)
$$

If $\lambda_0^* = 1$, then $DMU_o$ is an extreme efficient. Suppose there are $n$ extreme efficient DMUs in the sample, $DMU_j, j \in \mathbb{E}, j = 1, \ldots, n$.

Step 2: Determine all the FDEFs. The FDEF identification methods[2] can be found in Olesen and Petersen (1996) and Olesen and Petersen (2003).

Step 3: Determine and partition all the robust points by the algorithm *Determine and partition*. Then, we can get $\cup_{p \in \{1,\ldots,H\}} S_p^* = S^*$ meaning that DMUs with maximal FDEF participation in $S^*$ are divided into $H$ groups.

Step 4: Determine the Russell output measure based on the least-distance target. The following programming is solved using the DMUs of each group, $S_p^*, p = 1, \ldots, H$ as the reference set:

$$
\begin{aligned}
& Min\ W \sum_{r=1}^{S} (1 - z_r) + \frac{1}{s} \sum_{r=1}^{S} \frac{s_r^+ + s_r^-}{y_{ro}} \\
& s.t. \\
& \sum_{j \in J_p} \lambda_j x_{ij} \leqslant x_{io}, i = 1, \ldots, m \\
& \sum_{j \in J_p} \lambda_j y_{rj} = y_{ro} + s_r^+ - s_r^-, r = 1, \ldots, s \\
& s_r^+ - s_r^- = s_r, r = 1, \ldots, s \\
& s_r \geq -M(1 - z_r), r = 1, \ldots, s \\
& s_r \leq M z_r, r = 1, \ldots, s \\
& s_r^+ \geq 0, r = 1, \ldots, s \\
& s_r^- \geq 0, r = 1, \ldots, s \\
& z_r \in \{0, 1\} \\
& \lambda_j \geq 0.
\end{aligned} \quad (24)
$$

And the corresponding efficiency is:

---

[2] Olesen and Petersen (2003) offers two methods: MILP- and the Qhull-approach, for the identification of all FDEFs. We employ the MILP-approach in this study.

$$\theta_p(x_o, y_o) = \frac{1}{1 + \frac{1}{s}\sum_{r=1}^{s} \frac{s_r^{+*} + s_r^{-*}}{y_{ro}}}. \tag{25}$$

Then, we take the minimum value as the efficiency of the $DMU_o$:

$$\theta(x_o, y_o) = \underset{1 \leq p \leq H}{Min}\left(\theta_p(x_o, y_o)\right). \tag{26}$$

## 7. Empirical illustrations

In this study, "revenue" and "market price" do not refer to narrow commercial financial gains or monetary prices but rather to broader measures of output value and evaluation weights. For instance, the contributions of public universities in teaching, research, and societal services are challenging to quantify monetarily and are typically assessed through non-monetary evaluation weights established by governing bodies. However, the value of these outputs is susceptible to external uncertainties, such as policy changes that may alter the relative importance of talent cultivation or research. In this section, we utilize our proposed approach to study the China's 985 university project. The sample data comes from Zhu et al. (2022), and table 2 presents the specific details of the sample, which consists of two inputs: the number of researchers and size of the universities, and three outputs: National scientific awards (NSA), Scientific books, (SB) High-quality papers (HP) from 38 universities.

Table 2 Raw data on 38 universities in China.

| DMU |      | Input       |         | Output |    |      |
|-----|------|-------------|---------|--------|----|------|
|     |      | Researchers | Size    | NSA    | SB | HP   |
| 1   | PKU  | 4868        | 274.112 | 54     | 43 | 5414 |
| 2   | RUC  | 164         | 69.646  | 2      | 3  | 101  |
| 3   | TSU  | 4551        | 211.21  | 122    | 56 | 3526 |
| 4   | BUAA | 983         | 200     | 41     | 1  | 1939 |
| 5   | BIT  | 1756        | 289.467 | 28     | 26 | 1733 |
| 6   | CAU  | 1181        | 129.788 | 25     | 9  | 1447 |
| 7   | BNU  | 931         | 68.733  | 5      | 31 | 1169 |
| 8   | CUN  | 124         | 37.8    | 1      | 10 | 135  |
| 9   | NKU  | 1570        | 456.1   | 8      | 26 | 1808 |
| 10  | TU   | 1438        | 312     | 45     | 7  | 3116 |
| 11  | DUST | 1367        | 426.2   | 29     | 15 | 2171 |
| 12  | NEU  | 943         | 253     | 35     | 26 | 1377 |
| 13  | JLU  | 8501        | 611     | 57     | 15 | 1350 |
| 14  | HIT  | 2915        | 347.33  | 43     | 18 | 6164 |
| 15  | FDU  | 5509        | 244.32  | 36     | 33 | 5440 |
| 16  | TJU  | 2472        | 257     | 56     | 19 | 1716 |
| 17  | SJTU | 7110        | 326.2   | 114    | 43 | 6730 |
| 18  | ECNU | 953         | 207     | 9      | 12 | 1151 |
| 19  | NJU  | 1517        | 368.815 | 10     | 17 | 3366 |
| 20  | SEU  | 1458        | 392     | 9      | 38 | 2071 |
| 21 | ZJU | 3650 | 450.374 | 98 | 9 | 6185 |
| 22 | USTC | 1804 | 165 | 14 | 4 | 1964 |
| 23 | XMU | 435 | 600 | 18 | 5 | 1203 |
| 24 | SDU | 4433 | 533.333 | 62 | 14 | 2272 |
| 25 | OUC | 436 | 206.667 | 5 | 1 | 1666 |
| 26 | WHU | 2579 | 344.467 | 95 | 33 | 4058 |
| 27 | HUST | 3978 | 466.667 | 50 | 40 | 3636 |
| 28 | HNU | 1217 | 153.6 | 21 | 11 | 923 |
| 29 | CSU | 2999 | 392.4 | 83 | 19 | 1678 |
| 30 | SYSU | 5074 | 597.2 | 38 | 46 | 2851 |
| 31 | SCUT | 1776 | 294 | 33 | 15 | 1908 |
| 32 | CQU | 989 | 363.333 | 30 | 30 | 2142 |
| 33 | SCU | 4397 | 470 | 65 | 54 | 5174 |
| 34 | UESTC | 1721 | 333.333 | 22 | 22 | 2142 |
| 35 | XJTU | 2074 | 198.94 | 48 | 20 | 2756 |
| 36 | NPU | 1723 | 340 | 25 | 9 | 716 |
| 37 | NAFU | 1403 | 463.333 | 10 | 10 | 457 |
| 38 | LZU | 1096 | 253.8 | 24 | 3 | 1031 |

Table 3 The extreme efficient DMUs spanning each of the 15 FDEFs.

| DMU | FDEF | | | | | | | | | | | | | |
|---|---|---|---|---|---|---|---|---|---|---|---|---|---|---|
| | 1 | 2 | 3 | 4 | 5 | 6 | 7 | 8 | 9 | 10 | 11 | 12 | 13 | 14 |
| CQU | 1 | 1 | 1 | 1 | 1 | 1 | 1 | 1 | 0 | 0 | 0 | 0 | 0 | 0 |
| WHU | 1 | 1 | 0 | 0 | 1 | 1 | 0 | 0 | 1 | 1 | 0 | 1 | 1 | 0 |
| OUC | 1 | 1 | 1 | 1 | 0 | 0 | 0 | 0 | 1 | 1 | 1 | 0 | 0 | 0 |
| CUN | 1 | 0 | 1 | 0 | 1 | 0 | 1 | 0 | 1 | 0 | 1 | 1 | 0 | 0 |
| HIT | 0 | 0 | 0 | 0 | 0 | 0 | 0 | 0 | 1 | 1 | 1 | 1 | 1 | 1 |
| BUAA | 0 | 1 | 0 | 1 | 0 | 1 | 0 | 1 | 0 | 1 | 0 | 0 | 0 | 0 |
| XMU | 0 | 0 | 1 | 1 | 0 | 0 | 1 | 1 | 0 | 0 | 0 | 0 | 0 | 0 |
| BNU | 0 | 0 | 0 | 0 | 0 | 0 | 0 | 0 | 0 | 0 | 1 | 1 | 1 | 1 |
| NEU | 0 | 0 | 0 | 0 | 1 | 1 | 1 | 1 | 0 | 0 | 0 | 0 | 0 | 0 |
| SJTU | 0 | 0 | 0 | 0 | 0 | 0 | 0 | 0 | 0 | 0 | 0 | 0 | 1 | 1 |
| FDU | 0 | 0 | 0 | 0 | 0 | 0 | 0 | 0 | 0 | 0 | 0 | 0 | 0 | 1 |

First, we identify the extreme efficient DMUs according to the first step of the procedure for measuring robust efficiency based closest targets and find 11 extreme efficient universities, CQU, OUC, WHU, CUN, BUAA, HIT, XMU, BNU, NEU, SJTU, and FDU. Second, we determine 14 FDEFs by methods in Olesen and Petersen (1996) and Olesen and Petersen (2003). Table 3 shows which extreme efficient DMUs span each of the 14 FDEFs. Each column from the 2nd to the 14th in the table represents the DMUs that form a specific FDEF. The number 1 indicates that the DMU participates in forming the FDEF, while 0 means it does not. For example, the 2nd FDEF is spanned by CQU, OUC, WHU, and BUAA. Third, we determine and partition all the robust points by the algorithm *Determine and partition*. The detailed process is as follows:

Step 1. We obtain the set of FDEF each DMU belongs to by the loop. The results are shown in Table 3. For DMU $d$, the set of FDEFs corresponding to the columns where the element in its row is one is $K(d)$.

Step 2 and Step3. Compute the FDEF participation counts of each extreme efficient DMU and find that WHU and CQU have the highest participation counts, which is eight. Hence, we have $Maxcount = 8$ and $S^* = \{WHU, CQU\}$.

Step 4. We find that the eight FDEFs that WHU is involved with and the eight FDEFs that CQU is involved with are not entirely the same, so they form two teams. Hence, we have $GroupsCount = H = 2$. $S_1^* = \{CQU\}$ and the corresponding $G_1 = \{FDEF_1, FDEF_2, FDEF_3, FDEF_4, FDEF_5, FDEF_6, FDEF_7, FDEF_8\}$. $S_2^* = \{WHU\}$ and the corresponding $G_2 = \{FDEF_1, FDEF_2, FDEF_5, FDEF_6, FDEF_9, FDEF_{10}, FDEF_{12}, FDEF_{13}\}$. $\bigcup_{p \in \{1,\dots,H\}} T(S_p^*) = \{(x,y)|(x,y) = \lambda_j(x_j, y_j), \lambda_j \geq 0, j \in \{WHU, CQU\}\}$ is the entire production frontier formed by the robust points for the sample.

Fourth, using $\bigcup_{p \in \{1,\dots,H\}} T(S_p^*)$ as the reference set, we run the model from the step 4 in the previous section to compute the robust efficiency of each DMU based on the close targets.

Table 4 shows the robust efficiency ($Eff_1$) of each university based on closest targets, and the corresponding optimal slacks of the three outputs through which the optimal output, the closest robust targets, can be determined. In addition, we have provided several other efficiency measures for comparative analysis. $Eff_2$ is based on the efficiency based on closest targets; however, the selection of the targets does not account for robustness[3], please refer to Aparicio and Pastor (2013). $Eff_3$ is calculated using the weighted Russell model[4] based on the farthest distance, please refer to Ruggiero and Bretschneider (1998) and Chen et al. (2015). Note the models for obtaining $Eff_1$ and $Eff_2$ both incorporate FDEF to account for the marginal rate of substitution, whereas the model for obtaining $Eff_3$ does not consider it. For the sake of convenience in narration, we will assign the three models obtaining $Eff_1$, $Eff_2$ and $Eff_3$ to Model-1, Model-2, and Model-3 respectively.

The first thing that needs to be highlighted is that excluding WHU and CQU, all other DMUs are deemed inefficient for Model-1. According to the third to fifth columns of table 4, many DMUs exhibit negative output slack, indicating their ineffectiveness in managing risk; DMUs with positive output slack are considered inefficient from a technical or production perspective. In addition, it should be noted that it is reasonable for a DMU to exhibit positive slack for one output while showing negative slack for others. Only WHU and CQU have all output slacks equal to 0. However, in Model-2 and Model-3, in addition to WHU and CQU, there are 10 DMUs—TSU, BUAA, BNU, CUN, NEU, HIT, FDU, SJTU, XMU, and OUC—whose efficiency is 1. Among them, except for TSU, all the others are extreme efficient DMUs used to construct FDEF. TSU is an efficient but not an extreme efficient DMU.

---

[3] The Russell output measure is based on closest targets and the EFPPS and its projection target can resist risk through the only MRS corresponding to the EFPPS where it is located. However, the method of this study does not take EFPPS but FDEF as the alternative projection, and the differences between them can be referred to Olesen and Petersen (1996).

[4] This is an output-oriented weighted Russell model in which the weight vector of outputs is $(1/3, 1/3, 1/3)$. In addition, we can just achieve inefficiency by the model. Hence, to compare with other measures, we obtain efficiency by the equation: $\theta(x_o, y_o) = \frac{1}{1 + \frac{1}{s}\sum_{r=1}^{s} \frac{sr^*}{y_{ro}}}$, which is according with Model-1 and Model-3.

We use a production possibility set diagram with one input and two outputs to illustrate the reasons for these differences. In Fig. 2, line g-A-B-C-l is the production frontier of Model-3, like the production frontiers in most DEA models. The line segment a-A-B-C-f is the production frontier corresponding to the extended facet production possibility of Model-2. A, B, and C are three efficient DMUs, among which only B is robust. Therefore, in Model-1, both A and C need to use B as a benchmark to measure their efficiencies, resulting in A having slack in $y_1$, while C has slack in both of its outputs. BUAA, BNU, CUN, NEU, HIT, FDU, SJTU, XMU, and OUC are not robust and consider WHU or CQU as their closest target in Model-1. Hence, the $Eff_1$ is less than 1 for them.

Second, $Eff_1$ is generally less than $Eff_2$, and certain phenomena support this inference. Specifically, in our cases, only a few DMUs, such as RUC, BIT, and DUST, exhibit the opposite trend. Although Model-1, compared to Model-2, incorporates the robustness of the projection target to address risks, which is equivalent to adding an additional constraint, Model-1 allows for negative slack. As a result, $Eff_1$ and $Eff_2$ do not have an absolute relationship. The relative magnitude between the two is determined by the specific region in which the evaluated DMU is situated within the production possibility set. For example, any point on the line segment b-A-B-C-e including the robust point B is the E's benchmark candidates to obtain its $Eff_2$. Hence, its $Eff_1$ with B as the only projection target is less than or equal to $Eff_2$. We can conclude that for any DMU situated within the rectangle h-B-k-O whose selectable projection range encompasses B, the ordering relationship between $Eff_1$ and $Eff_2$ consistently holds, $Eff_1 \leq Eff_2$. For DMUs situated in other regions, such as D and F, the relationship between $Eff_1$ and $Eff_2$ is not determined; $Eff_1$ may be greater than, less than, or equal to $Eff_2$. For example, the projection candidates of F in Model-2 are all the points on the segment c-d not containing the optimal projection target, B, in Model-1. Hence, it is not definitively determined which of the two indicators is larger or smaller.

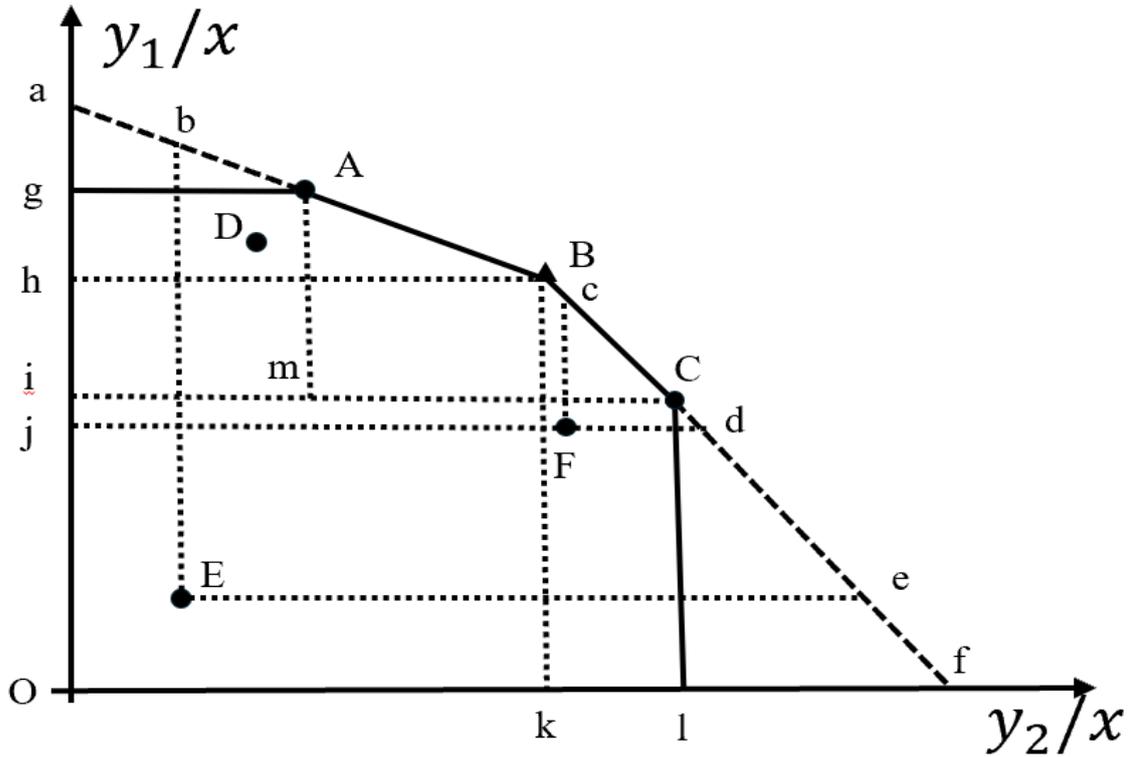

Fig. 2. The production possibility sets of Model-1, Model-2, and Model-3

Third, while $Eff_2$ is not less than $Eff_3$ in this sample, this does not imply that for every DMU, the $Eff_2$ and the $Eff_3$ consistently follow this size relationship. For DMUs in the region A-B-C-m, the projectable candidates of Model-2 and Model-3 are identical, with Model-2 aiming for minimal slack and Model-3 aiming for maximal slack. As a result, $Eff_2$ is not less than $Eff_3$ for these DMUs. However, the projectable segments of Model-2 and Model-3 differ for other DMUs. Specifically, Model-2's projectable segment includes the segments C-f and a-A, while Model-3's does not. Consequently, the $Eff_2$ and $Eff_3$ values for these DMUs do not always follow a fixed size relationship, potentially being greater than, equal to, or less than one another. Indeed, the analysis above suggests that $Eff_2$ being greater than or equal to $Eff_3$ is highly likely. In addition, $Eff_1$ and $Eff_3$ display a comparable relationship as well. In the region B-h-O-k, the projectable segments of Model-3 include B, meaning that for DMUs located in this region, $Eff_1$ is greater than or equal to $Eff_3$. The projectable segments of Model-1 and Model-3 differ for the DMUs not located in the region, which leads to $Eff_1$ and $Eff_3$ lacking a fixed order in their values, even though for most DMUs in this sample, $Eff_1$ is greater than $Eff_3$.

Table 4 Robus efficiency based on closest targets

| DMU | | Optimal slacks | | | Efficiency | | |
|---|---|---|---|---|---|---|---|
| | | NSA | SB | HP | Robust | Closest | Rusell |
| 1 | PKU | 0 | -24.2421 | -3107.3474 | 0.725 | 0.9964 | 0.9429 |
| 2 | RUC | 0.3645 | -2.1787 | 0 | 0.7676 | 0.4916 | 0.3891 |
| 3 | TSU | -63.7507 | -35.766 | -1037.8366 | 0.6733 | 1 | 1 |
| 4 | BUAA | -38.1212 | 0 | -1816.0303 | 0.6165 | 1 | 1 |
| 5 | BIT | -4.099 | -2.099 | -26.471 | 0.9252 | 0.7336 | 0.5784 |
| 6 | CAU | 8.8751 | 2.7671 | 0 | 0.8191 | 0.8735 | 0.5513 |
| 7 | BNU | 0 | -26 | -812 | 0.6618 | 1 | 1 |
| 8 | CUN | 0.8908 | -8.1092 | 0 | 0.6381 | 1 | 1 |
| 9 | NKU | 18 | 0 | 48.4 | 0.5685 | 0.6967 | 0.2931 |
| 10 | TU | 0 | 8.6316 | -1193.7895 | 0.6499 | 0.995 | 0.883 |
| 11 | DUST | 14.1818 | 0 | -326.4545 | 0.8243 | 0.811 | 0.4497 |
| 12 | NEU | -0.2637 | -13.9337 | 106.7898 | 0.8285 | 1 | 1 |
| 13 | JLU | -38.0924 | 3.9076 | 0 | 0.7636 | 0.2843 | 0.1121 |
| 14 | HIT | 8.8182 | 0 | -3950.5455 | 0.78 | 1 | 1 |
| 15 | FDU | 0 | -20.4947 | -3902.2316 | 0.6915 | 1 | 1 |
| 16 | TJU | 0 | 0.4526 | 676.0842 | 0.8778 | 0.7106 | 0.5126 |
| 17 | SJTU | -24.0378 | -11.75 | -2887.1947 | 0.7667 | 1 | 1 |
| 18 | ECNU | 7.1204 | 4.1204 | 0 | 0.7256 | 0.7507 | 0.4186 |
| 19 | NJU | 7 | 0 | -2152.2 | 0.6913 | 0.9495 | 0.4605 |
| 20 | SEU | 20.0056 | -8.9944 | 0 | 0.5495 | 0.8267 | 0.3701 |
| 21 | ZJU | -89 | 0 | -5542.4 | 0.6244 | 0.9648 | 0.4046 |
| 22 | USTC | 0 | 0.8632 | -1365.9789 | 0.767 | 0.8308 | 0.1662 |
| 23 | XMU | -1.9763 | 0.5661 | -518.537 | 0.821 | 1 | 1 |
| 24 | SDU | 0 | 7.5368 | 376.3789 | 0.8099 | 0.5123 | 0.203 |
| 25 | OUC | 0 | 0.7368 | -1452.4211 | 0.651 | 1 | 1 |
| 26 | WHU | 0 | 0 | 0 | 1 | 1 | 1 |
| 27 | HUST | -11.4678 | -1.4678 | -884.8029 | 0.8548 | 0.7134 | 0.5072 |
| 28 | HNU | -8.3174 | 1.6826 | -17.4638 | 0.8408 | 0.6431 | 0.4744 |
| 29 | CSU | -59.4986 | 4.5014 | 0 | 0.7588 | 0.7707 | 0.4232 |
| 30 | SYSU | 8 | 0 | 433.4 | 0.8922 | 0.5296 | 0.3481 |
| 31 | SCUT | 11.6673 | 0.516 | 0 | 0.8855 | 0.7443 | 0.4315 |
| 32 | CQU | 0 | 0 | 0 | 1 | 1 | 1 |
| 33 | SCU | 56.1262 | -11.9246 | 0 | 0.7345 | 0.8447 | 0.6497 |
| 34 | UESTC | 0 | 0 | -571.2 | 0.9184 | 0.7814 | 0.4932 |
| 35 | XJTU | 6.8653 | -0.9415 | -412.3836 | 0.8983 | 0.9705 | 0.7387 |
| 36 | NPU | 0.9091 | 0 | 390.7273 | 0.8375 | 0.5194 | 0.2194 |
| 37 | NAFU | 0 | 0 | 257 | 0.8421 | 0.3036 | 0.19 |
| 38 | LZU | 0.1363 | 5.3842 | 0 | 0.6249 | 0.6924 | 0.1477 |

## 8. Conclusion

In this study, we first propose a kind of robust benchmarking target, offering a novel approach for DMUs to navigate uncertainty without prior risk information. Each FDEF embeds a unique marginal substitution relationship, by which DMU can move to the position having the most revenue of the FDEF. Therefore, we identify the intersections of the maximum number of FDEFs as robust targets, as these points possess the greatest number of marginal substitution relationships, enabling them, in the event of a risk, to adjust their output with the highest probability to achieve globally optimal revenue. Furthermore, we develop a novel well-defined efficiency measure based on robust and closest targets. Finally, we demonstrate the application of the proposed measure, which is based on robust and closest targets, using a dataset comprising 38 universities from China's 985 Project. Through this empirical example, we reveal notable differences in the resulting targets when applying the criterion of robust and closest targets, as opposed to the separate criteria of the closest targets and the farthest targets. These findings emphasize the practical significance of identifying robust and closest targets in real-world applications

Future research could enhance the proposed robust benchmarking approach by evaluating its resilience across diverse risk scenarios, including extreme or non-linear disruptions, to confirm its efficacy beyond the current marginal substitution assumptions within the DEA framework. Furthermore, integrating stochastic modeling or cost-benefit analyses of real options could refine the methodology, strengthening its value as a comprehensive tool for DMUs navigating uncertainty without prior risk information. Additionally, incorporating the cost of production adjustment into the analytical framework could improve the practicality of the approach, by addressing the economic implications of adapting production under uncertain conditions.

**References**

Aigner, D., Lovell, C. A. K., & Schmidt, P. (1977). Formulation and estimation of stochastic frontier production function models. *Journal of Econometrics*, *6*(1), 21-37. https://doi.org/10.1016/0304-4076(77)90052-5

Aigner, L., & Asmild, M. (2023). Identifying the most important set of weights when modelling bad outputs with the weak disposability approach. *European Journal of Operational Research*, *310*(2), 751-759. https://doi.org/10.1016/j.ejor.2023.02.021

Amirteimoori, A., & Kordrostami, S. (2010). A Euclidean distance-based measure of efficiency in data envelopment analysis. *Optimization*, *59*(7), 985-996. https://doi.org/10.1080/02331930902878333

Ang, S., An, Q. X., Yang, F., & Ji, X. (2019). Target setting with minimum improving costs in data envelopment analysis: A mixed integer linear programming approach. *Expert Systems*, *36*(4), Article e12408. https://doi.org/10.1111/exsy.12408

Aparicio, J. (2016). A survey on measuring efficiency through the determination of the least distance in data envelopment analysis. *Journal of Centrum Cathedra*, *9*(2), 143-167. https://doi.org/10.1108/JCC-09-2016-0014

Aparicio, J., & Pastor, J. T. (2013). A well-defined efficiency measure for dealing with closest targets in DEA. *Applied Mathematics and Computation*, *219*(17), 9142-9154. https://doi.org/10.1016/j.amc.2013.03.042

Aparicio, J., & Pastor, J. T. (2014). Closest targets and strong monotonicity on the strongly efficient frontier in DEA. *Omega*, *44*, 51-57. https://doi.org/10.1016/j.omega.2013.10.001

Aparicio, J., Ruiz, J. L., & Sirvent, I. (2007). Closest targets and minimum distance to the Pareto-efficient frontier in DEA. *Journal of Productivity Analysis*, *28*(3), 209-218. https://doi.org/10.1007/s11123-007-0039-5

Arabmaldar, A., Hatami-Marbini, A., Loske, D., Hammerschmidt, M., & Klumpp, M. (2024). Robust data envelopment analysis with variable budgeted uncertainty. *European Journal of Operational Research*, *315*(2), 626-641. https://doi.org/10.1016/j.ejor.2023.11.043

Baek, C., & Lee, J.-d. (2009). The relevance of DEA benchmarking information and the Least-Distance Measure. *Mathematical and Computer Modelling*, *49*(1), 265-275. https://doi.org/10.1016/j.mcm.2008.08.007

Banker, R. D., Charnes, A., & Cooper, W. W. (1984). Some Models for Estimating Technical and Scale Inefficiencies in Data Envelopment Analysis. *Management Science*, *30*(9), 1078-1092. https://doi.org/10.1287/mnsc.30.9.1078

Bertsimas, D., & Sim, M. (2004). The Price of Robustness. *Operations Research*, *52*(1), 35-53. https://doi.org/10.1287/opre.1030.0065

Bougnol, M. L., Dulá, J. H., & Rouse, P. (2012). Interior point methods in DEA to determine non-zero multiplier weights. *Computers & Operations Research*, *39*(3), 698-708. https://doi.org/10.1016/j.cor.2011.05.006

Camanho, A. S., & Dyson, R. G. (2005). Cost efficiency measurement with price uncertainty: a DEA application to bank branch assessments. *European Journal of Operational Research*, *161*(2), 432-446. https://doi.org/10.1016/j.ejor.2003.07.018

Charnes, A., Cooper, W. W., & Rhodes, E. (1978). Measuring the efficiency of decision making units. *European Journal of Operational Research*, *2*(6), 429-444. https://doi.org/10.1016/0377-2217(78)90138-8

Chen, K., & Zhu, J. (2019). Computational tractability of chance constrained data envelopment analysis. *European Journal of Operational Research*, *274*(3), 1037-1046. https://doi.org/10.1016/j.ejor.2018.10.039

Chen, P., Yu, M., Chang, C., Hsu, S., & Managi, S. (2015). Nonradial Directional Performance Measurement with Undesirable Outputs: An Application to OECD and Non-OECD Countries. *International Journal of Information Technology & Decision Making*, *14*(03), 481-520. https://doi.org/10.1142/S0219622015500091

Chen, Y., Wen, M., & Wang, F. (2019). A new uncertain DEA model for evaluation of scientific research personnel. *Journal of Intelligent & Fuzzy Systems*, *37*(4), 5633-5640. https://doi.org/10.3233/jifs-190784

Chung, Y. H., Färe, R., & Grosskopf, S. (1997). Productivity and Undesirable Outputs: A Directional Distance Function Approach. *Journal of Environmental Management*, *51*(3), 229-240. https://doi.org/10.1006/jema.1997.0146

Cooper, W. W., Park, K. S., & Yu, G. (2001). An illustrative application of idea (imprecise Data Envelopment Analysis) to a Korean mobile telecommunication company. *Operations Research*, *49*(6), 807-820. https://doi.org/10.1287/opre.49.6.807.10022

Cooper, W. W., Ruiz, J. L., & Sirvent, I. (2007). Choosing weights from alternative optimal solutions of dual multiplier models in DEA. *European Journal of Operational Research*, *180*(1), 443-458. https://doi.org/10.1016/j.ejor.2006.02.037

Ehrgott, M., Holder, A., & Nohadani, O. (2018). Uncertain Data Envelopment Analysis. *European Journal of Operational Research*, *268*(1), 231-242. https://doi.org/10.1016/j.ejor.2018.01.005

Emrouznejad, A., Amin, G. R., Ghiyasi, M., & Michali, M. (2023). A review of inverse data envelopment analysis: origins, development and future directions. *Ima Journal of Management Mathematics*, *34*(3), 421-440. https://doi.org/10.1093/imaman/dpad006

Emrouznejad, A., Podinovski, V., Charles, V., Lu, C., & Moradi-Motlagh, A. (2025). Rajiv Banker's lasting impact on data envelopment analysis. *Annals of Operations Research*. https://doi.org/10.1007/s10479-025-06473-3

Färe, R., Grosskopf, S., & Weber, W. L. (2006). Shadow prices and pollution costs in U.S. agriculture. *Ecological Economics*, *56*(1), 89-103. https://doi.org/10.1016/j.ecolecon.2004.12.022

Fare, R., He, X., Li, S., & Zelenyuk, V. (2019). A unifying Framework for Farrell Profit Efficiency Measurement. *Operations Research*, *67*(1), 183-197. https://doi.org/10.1287/opre.2018.1770

Fukuyama, H., Maeda, Y., Sekitani, K., & Shi, J. (2014). Input–output substitutability and strongly monotonic p-norm least distance DEA measures. *European Journal of Operational Research*, *237*(3), 997-1007. https://doi.org/10.1016/j.ejor.2014.02.033

Godin, F., Van Son, L., & Trottier, D.-A. (2019). Option pricing under regime-switching models: Novel approaches removing path-dependence. *Insurance Mathematics & Economics*, *87*, 130-142. https://doi.org/10.1016/j.insmatheco.2019.04.006

González, E., & Álvarez, A. (2001). From efficiency measurement to efficiency improvement: The choice of a relevant benchmark. *European Journal of Operational Research*, *133*(3), 512-520. https://doi.org/10.1016/S0377-

2217(00)00195-8

Guo, F., Gregory, J., & Kirchain, R. (2020). Incorporating cost uncertainty and path dependence into treatment selection for pavement networks. *Transportation Research Part C-Emerging Technologies*, *110*, 40-55. https://doi.org/10.1016/j.trc.2019.11.015

Hasannasab, M., Margaritis, D., Roshdi, I., & Rouse, P. (2024). Positive weights in data envelopment analysis. *Journal of Productivity Analysis*, *62*(3), 351-363. https://doi.org/10.1007/s11123-023-00682-3

Huang, K.-M., Kuo, I. D., & Wang, R.-T. (2022). Resale options and heterogeneous beliefs. *Journal of Futures Markets*, *42*(6), 1067-1083. https://doi.org/10.1002/fut.22319

Lertworasirkul, S., Fang, S. C., Joines, J. A., & Nuttle, H. L. W. (2003). Fuzzy data envelopment analysis (DEA): a possibility approach. *Fuzzy Sets and Systems*, *139*(2), 379-394, Article Pii s0165-0114(02)00484-0. https://doi.org/10.1016/s0165-0114(02)00484-0

Mergoni, A., Emrouznejad, A., & De Witte, K. (2024). Fifty years of Data Envelopment Analysis. *European Journal of Operational Research*. https://doi.org/10.1016/j.ejor.2024.12.049

Olesen, O. B., & Petersen, N. C. (1996). Indicators of Ill-Conditioned Data Sets and Model Misspecification in Data Envelopment Analysis: An Extended Facet Approach. *Management Science*, *42*(2), 205-219. https://doi.org/10.1287/mnsc.42.2.205

Olesen, O. B., & Petersen, N. C. (2003). Identification and Use of Efficient Faces and Facets in DEA. *Journal of Productivity Analysis*, *20*(3), 323-360. https://doi.org/10.1023/A:1027303901017

Olesen, O. B., & Petersen, N. C. (2024). Facet analysis in data envelopment analysis: some pitfalls of the CRS models. *Journal of Productivity Analysis*. https://doi.org/10.1007/s11123-023-00715-x

Pan, Y., Wu, J., Zhang, C.-C., & Nasir, M. A. (2024). Measuring carbon emission performance in China's energy market: Evidence from improved non-radial directional distance function data envelopment analysis. *European Journal of Operational Research*. https://doi.org/10.1016/j.ejor.2024.11.019

Petersen, N. C. (2018). Directional Distance Functions in DEA with Optimal Endogenous Directions. *Operations Research*, *66*(4), 1068-1085. https://doi.org/10.1287/opre.2017.1711

Petridis, K., Dey, P. K., & Emrouznejad, A. (2017). A branch and efficiency algorithm for the optimal design of supply chain networks. *Annals of Operations Research*, *253*(1), 545-571. https://doi.org/10.1007/s10479-016-2268-3

Portela, M. C. A. S., & Thanassoulis, E. (2006). Zero weights and non-zero slacks: Different solutions to the same problem. *Annals of Operations Research*, *145*(1), 129-147. https://doi.org/10.1007/s10479-006-0029-4

Ruggiero, J., & Bretschneider, S. (1998). The weighted Russell measure of technical efficiency. *European Journal of Operational Research*, *108*(2), 438-451. https://doi.org/10.1016/S0377-2217(97)00150-1

Schaffnit, C., Rosen, D., & Paradi, J. C. (1997). Best practice analysis of bank branches: An application of DEA in a large Canadian bank. *European Journal of Operational Research*, *98*(2), 269-289. https://doi.org/10.1016/S0377-2217(96)00347-5

Simar, L., & Wilson, P. W. (1998). Sensitivity Analysis of Efficiency Scores: How to Bootstrap in Nonparametric Frontier Models. *Management Science*, *44*(1), 49-61. http://www.jstor.org/stable/2634426

Thompson, R. G., Dharmapala, P. S., Humphrey, D. B., Taylor, W. M., & Thrall, R. M. (1996). Computing DEA/AR efficiency and profit ratio measures with an illustrative bank application. *Annals of Operations Research*, *68*, 301-327. https://doi.org/10.1007/BF02207220

Tone, K. (2002). A slacks-based measure of super-efficiency in data envelopment analysis. *European Journal of Operational Research*, *143*(1), 32-41. https://doi.org/10.1016/s0377-2217(01)00324-1

Tone, K., Chang, T.-S., & Wu, C.-H. (2020). Handling negative data in slacks-based measure data envelopment analysis models. *European Journal of Operational Research*, *282*(3), 926-935. https://doi.org/10.1016/j.ejor.2019.09.055

Topcu, T. G., & Triantis, K. (2022). An ex-ante DEA method for representing contextual uncertainties and stakeholder risk preferences. *Annals of Operations Research*, *309*(1), 395-423. https://doi.org/10.1007/s10479-021-

04271-1

Zhu, J. (2022). DEA under big data: data enabled analytics and network data envelopment analysis. *Annals of Operations Research*, *309*(2), 761-783. https://doi.org/10.1007/s10479-020-03668-8

Zhu, Q., Aparicio, J., Li, F., Wu, J., & Kou, G. (2022). Determining closest targets on the extended facet production possibility set in data envelopment analysis: Modeling and computational aspects. *European Journal of Operational Research*, *296*(3), 927-939. https://doi.org/10.1016/j.ejor.2021.04.019